\documentclass[prd,aps,preprint,tightenlines,nofootinbib]{revtex4}

\pdfoutput=1

 \usepackage[dvips,final]{graphicx}
  \usepackage{amssymb}
   \usepackage{amsmath}
    \usepackage{amsfonts}
     \usepackage{epsfig}
      \usepackage{bm} 
      \usepackage[dvipsnames,usenames]{color}
      \usepackage{comment}
      \usepackage{float}
      
      \usepackage[utf8]{inputenc}

\begin{document}

\title{DPS mechanism for associated $c\bar c b\bar b$ production in $AA$ UPCs}

\author{Edgar Huayra$^{1}$}
\email{yuberth022@gmail.com}

\author{Emmanuel G. de Oliveira$^{1}$}
\email{emmanuel.de.oliveira@ufsc.br}

\author{Roman Pasechnik$^{1,2,3}$}
\email{Roman.Pasechnik@thep.lu.se}

\affiliation{
\\
{$^1$\sl Departamento de F\'isica, CFM, Universidade Federal 
de Santa Catarina, C.P. 476, CEP 88.040-900, Florian\'opolis, 
SC, Brazil
}\\
{$^2$\sl
Department of Astronomy and Theoretical Physics, Lund
University, SE-223 62 Lund, Sweden
}\\
{$^3$\sl Nuclear Physics Institute ASCR, 25068 \v{R}e\v{z}, 
Czech Republic\vspace{1.0cm}
}}

\begin{abstract}
\vspace{0.5cm}
We discuss the associated $c\bar{c}$ and $b\bar{b}$ quark pairs
production in the double-parton scattering (DPS) process in ultraperipheral (UPCs)
$AA$ collisions. We derive an analogue of the inclusive DPS pocket formula and the 
photon-energy dependent effective cross section considering an overlap between 
the hard SPS scatterings. We provide numerical predictions for the DPS cross sections for
the $c\bar{c}b\bar{b}$ production process at the typical energies of $AA$ UPCs 
at the LHC and FCC colliders and also characterize the $A$ dependence of 
the total UPC DPS cross section.
\end{abstract}

\pacs{12.38.-t,12.38.Lg,12.39.St,13.60.-r,13.85.-t}

\maketitle

\section{Introduction}
\label{Sect:intro}

The double- or, generally, multi-parton interaction (MPI) processes become increasingly important in hadron-hadron and especially nucleus-nucleus collisions at high energies. In particular, the ratio of probabilities of the double-parton scattering (DPS) to single-parton scattering (SPS) grows in energy \cite{Paver:1982yp,Mekhfi:1983az,Sjostrand:1987su} and cannot be neglected already at the LHC. In connection to the LHC measurements, a significant amount of work, both experimental and theoretical, has been done in analysis of MPI contributions to hadroproduction reactions with many different final states, see e.g.~Refs.~\cite{Gaunt:2009re,Diehl:2011yj,Manohar:2012jr,Aaij:2015wpa,Blok:2016lmd,Rinaldi:2015cya,Maciula:2018mig,Abe:1993rv,Aaboud:2018tiq} and references therein. Most common examples with a significant DPS effect include meson pairs \cite{Maciula:2018mig}, four identified jets \cite{Abe:1993rv} or leptons \cite{Aaboud:2018tiq} etc. In fact, the MPIs are naturally accounted for in most commonly used event generators \cite{Sjostrand:1987su,Fedkevych:2019ofc}.

From theoretical viewpoint, the DPS processes are connected to yet poorly known double-parton 
distribution functions (dPDFs) which are of a predominantly non-perturbative QCD origin 
and remain highly uncertain (for a detailed review on the theoretical basis of dPDFs, 
see e.g.~Ref.~\cite{Diehl:2017wew} and references therein). These objects represent
number densities of correlated pairs of two colored partons (quark and gluons) 
in the proton or a nucleus carrying longitudinal momentum fractions $x_1$ and $x_2$ 
that are found at a particular transverse relative separation ${\bf b}$ 
between subsequent two hard scatterings \cite{Calucci:1999yz}. While the first-principle 
calculations of dPDFs with non-perturbative parton correlations are yet not feasible 
theoretically, several models attempting to capture the most relevant dPDFs properties 
are being advised in the literature, see e.g.~Refs.~\cite{Chang:2012nw,Rinaldi:2013vpa,
Rinaldi:2014ddl,Rinaldi:2015cya,Blok:2016lmd}. 

Relevant phenomenological information about dynamics of dPDFs is contained in a measurable quantity 
known as effective cross section, $\sigma_{\rm eff}$ that is being extensively studied in various 
particle production channels~\cite{Akesson:1986iv,Alitti:1991rd,Abe:1993rv,Aaij:2012dz,Abazov:2009gc,
Aad:2014kba,Aaij:2016bqq,Sirunyan:2017hlu,Aaboud:2018tiq}). 
Among the final states, such processes as the associated double open heavy flavor 
production and heavy flavor production in association with jets are typically 
considered as a powerful toolkit for probing the DPS dynamics, see its earlier 
studies in non-peripheral collisions in Refs.~\cite{Aaij:2012dz,Luszczak:2011zp,
DelFabbro:2002pw,Cazaroto:2013fua,Maciula:2018mig,Maciula:2017egq}

In our earlier analysis of Ref.~\cite{Huayra:2019iun} we performed a first study of double 
open heavy flavor production, specifically, in high-energy ultraperipheral collisions (UPCs) 
focusing on the  $c\bar c b \bar b$ final states produced off the proton target. An enhanced 
double-photon flux coming from the incident nucleon has been considered as a probe for 
the double-gluon density in the proton. Among the advantages of using the $pA$ and $AA$ UPCs compared 
to more standard central $pp$ collisions for probing the MPIs are a relatively low QCD background provided 
by tagging on the final-state nucleus and a strongly enhanced quasi-real Weisz\"acker-Williams (WW) photon 
flux \cite{vonWeizsacker:1934nji,Williams:1934ad} in the incident nucleus, with a rather broad photon spectrum.

As usual, the effective cross section in this case corresponds to an effective transverse overlap area 
between hard single-parton scatterings and is conventionally defined as a ratio between the product 
of SPS cross sections, $\sigma^{c\bar c}_{\rm SPS}$ and $\sigma^{b\bar b}_{\rm SPS}$, over the production 
cross section going via the DPS mechanism, $\sigma^{c\bar c  b\bar b}_{\rm DPS}$, such that
\begin{eqnarray}
\sigma_{\rm eff} = \frac{\sigma^{c\bar c}_{\rm SPS}\sigma^{b\bar b}_{\rm SPS}}
{\sigma^{c\bar c  b\bar b}_{\rm DPS}} \,,
\label{eq:pocket}
\end{eqnarray}
Under an approximation when the effective cross section is not dependent on the kinematics of the underlined 
process and is considered as a constant geometrical factor, Eq.~(\ref{eq:pocket}) is known as the ``pocket formula''. 
Switching from the two-gluon initial state to the two-photon one coming from the initial nucleus, it was shown 
in Ref.~\cite{Huayra:2019iun} that the effective cross section becomes significantly larger than 
for four-parton initial state and amounts to, roughly, a few dozens of barns. Besides, due to a wide-spread 
photon distribution, $\sigma_{\rm eff}$ strongly depends on the photon longitudinal momentum fraction such 
that a naive multiplication of SPS cross sections in Eq.~(\ref{eq:pocket}) does not apply any longer and 
one has to perform a convolution in the photon momentum fraction.

In the current study, we consider the same type of process as an efficient probe for the double-gluon 
density in the target nucleus at small-$x$ and compute the corresponding DPS contribution to the observables 
at different energies. For this purpose, we extend our previous study \cite{Huayra:2019iun} to heavy-ion $AA$ 
UPCs with the focus of probing the gluon dPDFs in the $A+A \to A + (c\bar c b \bar b) + X$ reaction 
by means of double-photon exchange with another nucleus at large impact parameters. Particularly, this process represents 
a novel and clean way of probing small-$x$ gluon dPDFs in a nucleus competitive to more standard processes 
in central collisions.

The paper is organised as follows. In Sec.~\ref{Sect:ccbb-SPS}, we discuss the basics of the SPS contribution to 
the inclusive differential $c\bar c b \bar b$ production cross section setting up all the necessary ingredients for 
the DPS analysis. In Sec.~\ref{Sect:ccbb-DPS}, we describe two distinct contributions to the DPS mechanism of 
$c\bar c b \bar b$ production in $AA$ UPCs coming from uncorrelated (on two different nucleons) 
and correlated (on a single nucleon) SPS processes. Here, we also derive the generalised UPC pocket formula
accounting for both contributions and the corresponding effective cross section. In Sec.~\ref{Sect:Results},
we present numerical results for the effective cross section and for the differential DPS  cross sections
for double heavy-quark pair production in $AA$ UPCs at typical LHC and FCC energies. Finally, basic conclusions
are made in Sect.~\ref{Sect:Conclusions}.

\section{Single-parton scattering}
\label{Sect:ccbb-SPS}

The kinematics of the considered $A+A \to A + (c\bar c b \bar b) + X$ process and details of calculations are rather 
similar to those in the $A+p \to A + (c\bar c b \bar b) + X$ previously discussed in Ref.~\cite{Huayra:2019iun}. 
So, here we only briefly discuss  the most crucial elements of the formalism particularly relevant for the 
direct $c\bar c b \bar b$ production reaction in $AA$ UPCs.

To start with let us define the photon and gluon longitudinal momentum fractions as follows
\begin{eqnarray}
    \xi_i = \frac{m_{i,\perp}}{\sqrt{s}} ( e^{y_{Q_i}} + e^{y_{\bar Q_i} } ) \,, \qquad 
    x_i = \frac{m_{i,\perp}}{\sqrt{s}} ( e^{-y_{Q_i}} + e^{-y_{\bar Q_i} } ) \,, \qquad 
    m^2_{i,\perp} = m_{Q_i}^2 + p_{i,\perp}^2 \,,
    \label{kinematics}
\end{eqnarray}
Here, $i=1,2$ correspond to separate hard SPS photon-gluon fusion $\gamma g\to c\bar c$ and 
$\gamma g\to b\bar b$ subprocesses, respectively, such that $Q_{1,2}\equiv c,b$, while $m_{i,\perp}$ and $y_{Q_i}$ 
($y_{\bar Q_i}$) are the (anti)quark transverse mass and rapidity, respectively.

The cross section for $Q\bar Q$-pair production in the SPS $AA$ UPCs can be represented
as a convolution of parton-level $\gamma+g \to Q\bar{Q}$ subprocess cross section off 
a given nucleon in the nucleus target with the spatial nucleon density in the nucleus given 
by the so-called nuclear thickness function $\rho(\vec b_p)$, as well as the gluon distribution in 
the nucleon, $G_g^A$, in the following form\footnote{We assume for simplicity that the scattered 
nuclei are the same.}
\begin{eqnarray}
\nonumber
\frac{d^2\sigma_{AA \rightarrow XA + Q\bar{Q}}}{dy_{Q}dy_{\bar{Q}}} 
& = & \int d^2 \vec{b} d^2 \vec{b}_\gamma d^2 \vec{b}_p d^2 \vec{b}_g d \xi d x\, 
\Theta(b - 2R_A) \delta^{(2)} (\vec{b} + \vec{b}_p + \vec{b}_g - \vec{b}_\gamma) \\
& \times & 
\,N_\gamma(\xi,\vec{b}_\gamma) G^{\rm A}_g(x, \vec{b}_g) \rho(\vec b_p)
\frac{d^2\hat{\sigma}_{\gamma g \rightarrow Q\bar{Q}}}{dy_{Q}dy_{\bar{Q}}} \, , \\
\frac{d^2\hat{\sigma}_{\gamma g \rightarrow Q\bar{Q}}}{dy_{Q}dy_{\bar{Q}}} \,
&=& \int d\hat{t} \frac{d\hat{\sigma}_{\gamma g \rightarrow Q\bar{Q}}}{d\hat{t}}
 \, \delta\left (y_{Q} - \frac12 \ln \left( \frac{\xi}{x} \frac{\hat{u}}{\hat{t}} \right) \right) 
 \, \delta\left (y_{\bar{Q}} - \frac12 \ln \left( \frac{\xi}{x} \frac{\hat{t}}{\hat{u}} \right) \right) \,, \\
&& \hat{t}= (p_Q - p_\gamma)^2 - m^2_Q = - \sqrt{\hat{s}}
\bigg(\frac{\sqrt{\hat{s}}}{2}-
\sqrt{\frac{\hat{s}}{4}-m^2_Q-p^2_\perp} \bigg)\,,
\label{eq:SPS}
\end{eqnarray}
in terms of the $Q\bar Q$ invariant mass squared $\hat{s}$, 
nucleus radius $R_A$, (quasi-real) WW single-photon flux in the nucleus $N_\gamma$,
\begin{eqnarray}
\frac{d^3 N_\gamma(\xi,{\vec b})}{d\xi d^2{\vec b}}= 
\frac{\sqrt{s}}{2}\frac{Z^2\alpha k^2}{\pi^2\omega b^2} \Big[ K^2_1(k)+\frac{1}
{\gamma^2}K^2_0(k) \Big]\,, \qquad k=\frac{b\, \omega}{\gamma} \,, \qquad 
\xi = \frac{2 \omega}{\sqrt{s}} \,,
\end{eqnarray}
and the nucleon single-gluon $G^{\rm A}_g$ distribution \cite{Frankfurt:2010ea},
\begin{align}\nonumber
G^{\rm A}_g(x, {\vec b}_g) = A\, g_A(x)\, f_g({\vec b}_g) \,, \\
f_g ({\vec b}) = \frac{\Lambda^2}{2 \pi} \frac{\Lambda  b}{2} K_1(\Lambda  b) \,, 
\quad \int d^2 {\vec b} \, f_g ({\vec b}) = 1 \,.
\label{Gg}
\end{align}
Above, $A$ and  $Z$ are the atomic mass and the electric charge of the nucleus, respectively,
$K_{0,1}$ are the modified Bessel functions of the second kind, $\alpha$ is the fine 
structure constant, $\omega$ is the exchanged photon energy, $\gamma = \sqrt{s} / 2 m_p$ 
is the Lorentz factor defined in terms of the total center-of-mass (c.m.) energy per 
nucleon squared, $s$, and the proton mass $m_p=0.938$ GeV. The distribution $g_A(x)$ is the collinear 
gluon density in the nucleon in a given nucleus $A$ for which the EPPS16nlo~\cite{Eskola:2016oht} parameterisation 
has been used (with the factorisation scale $\mu_F = \hat{s}$), $f_g({\vec b})$ 
is the normalised spatial gluon density in a nucleon inside the nucleus, the scale 
parameter $\Lambda \approx 1.5$ GeV. In what follows, the thickness function $\rho(\vec b_p)$ 
has been adopted in the Woods-Saxon parameterisation \cite{Woods:1954zz},
\begin{align}
\rho(\vec b_p) = \rho_0\, \int dz\, \frac{1}{1+\exp\Big(\frac{\sqrt{\vec{b}_p^2 + z^2} - R_A}{\delta}\Big)} \,, 
\qquad \int d^2 {\vec b_p} \, \rho ({\vec b}_p) = 1 \,,
\end{align}
where $\rho_0$ is an overall normalisation, and $\delta = 0.459$ fm with $R_A = [1.1 A^{1/3} - 0.65 A^{-1/3}]$ fm to $4 \leq A \leq 208$. Provided that the spatial gluon
distribution in the nucleon is much narrower than that of the photon, the SPS cross section (\ref{eq:SPS}) 
can be conveniently rewritten in terms of the collinear gluon density as follows
\begin{eqnarray}
\frac{d^2\sigma_{AA \rightarrow A(Q\bar{Q})X}}{dy_{Q}dy_{\bar{Q}}} 
= \int d \xi \int d x \, 
\overline{N}_\gamma(\xi)\, A g_A(x)
\frac{d^2\hat{\sigma}_{\gamma g \rightarrow Q\bar{Q}}}{dy_{Q}dy_{\bar{Q}}} \,
 \int d^2 \vec{b}\, \Theta(b - 2R_A) T_{g\gamma} (\xi, \vec{b}) \,.
\label{eq:SPS-b}
\end{eqnarray}
Here, the number distribution of interacting photons averaged over 
their transverse positions outside the nucleus target reads
\begin{align}
\label{eq:Nbar}
\overline{N}_\gamma(\xi) = \int d^2 b \, \Theta(b - R_A) N_\gamma (\xi,\vec b) \,,
\end{align}
and the overlap function
\begin{eqnarray}
\label{eq:Tg}
T_{g\gamma} (\xi, \vec{b}) = \frac{1}{\overline{N}_\gamma(\xi)} \int d^2 \vec{b}_p \, 
\int d^2 \vec{b}_\gamma  \, 
\Theta(b_\gamma - R_A) N_\gamma(\xi,\vec{b}_\gamma) f_g(\vec{b}_\gamma - \vec{b_p} - \vec b ) \, 
\rho(\vec b_p) \,,
\end{eqnarray}
provides an information about the impact parameter dependence of the matrix element squared.
\begin{figure*}[!htbt]
\begin{minipage}{0.48\textwidth}
 \centerline{\includegraphics[width=1.0\textwidth]{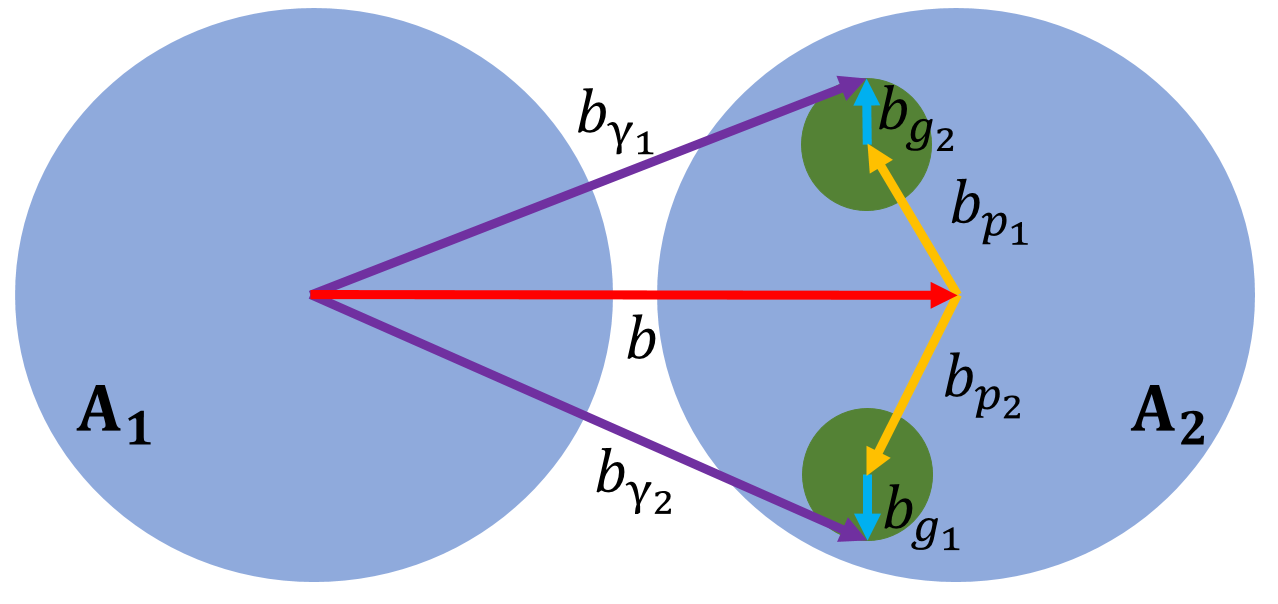}}
\end{minipage} \hfill
\begin{minipage}{0.48\textwidth}
 \centerline{\includegraphics[width=1.0\textwidth]{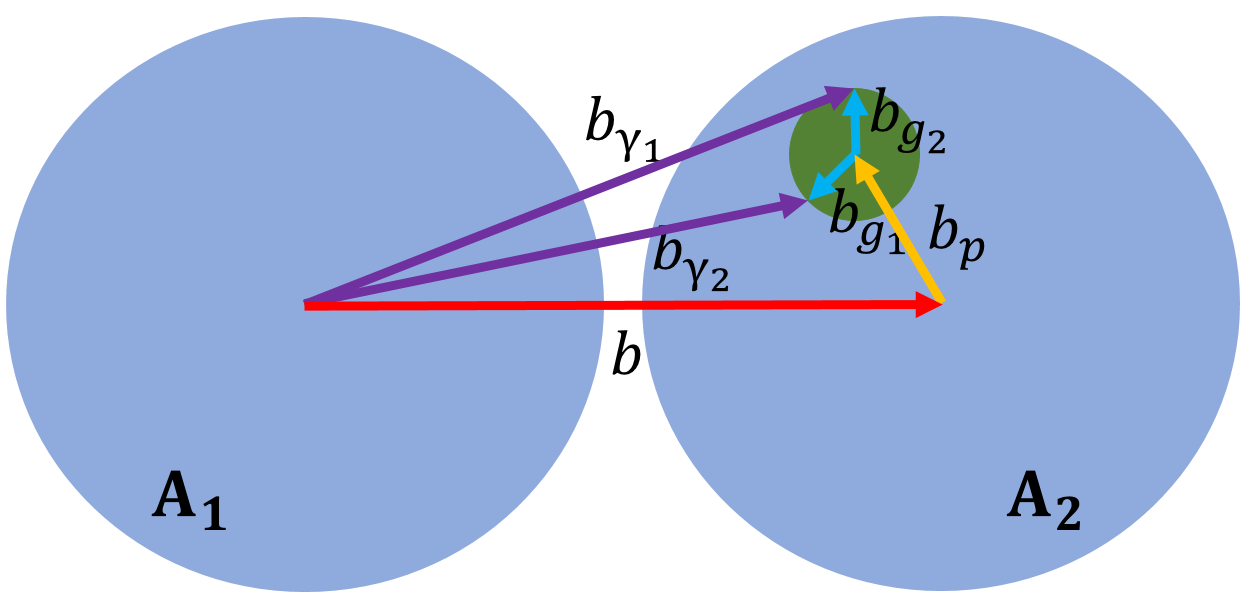}}
\end{minipage} \hfill
\caption{Two distinct contributions to the DPS cross section in $AA$ UPCs. On the left, the two photons from the projectile interact with two gluons originating from different nucleons; its contribution to the DPS cross section is proportional to $A-1$ and will be the dominant one for large $A$. On the right, the photons interact with gluons from the same nucleon.}
\label{fig:two-contributions}
\end{figure*}

\section{Double-parton scattering}
\label{Sect:ccbb-DPS}

In variance to the proton target case studied earlier in Ref.~\cite{Huayra:2019iun} and generalizing 
the $pp$ approach of the Ref~\cite{Calucci:1999yz}, for the nucleus target there are two distinct
contributions to the DPS UPC cross section
\begin{eqnarray}
    \sigma^{\rm DPS}_{\rm AA} = \sigma^{\rm DPS}_{\rm I} + \sigma^{\rm DPS}_{\rm II} \,.
\end{eqnarray}
The first contribution here corresponds to the situation when two SPS gluons are taken from 
two different nucleons in the nucleus as illustrated in Fig.~\ref{fig:two-contributions} 
(left), while the second contribution emerges when the two gluons come from a single nucleon 
as in Fig.~\ref{fig:two-contributions} (right). In what follows, we neglect any differences between 
proton and neutron in the nucleus target which is an approximation suitable in the high-energy 
regime.

\subsection{Scattering off different nucleons}
\label{Sect:case-I}

The DPS cross section for the first case in Fig.~\ref{fig:two-contributions} 
(left) can be represented as a convolution of the parton-level SPS cross 
sections \cite{Huayra:2019iun},
\begin{eqnarray}
\nonumber
&& \frac{d^4\sigma^{\rm DPS}_{\rm I}}{dy_{c}dy_{\bar{c}}dy_{b}dy_{\bar{b}}}
= \int d^2 b \, \Theta(b - 2R_A) \int d^2 \vec{b}_{\gamma,1} \, \Theta (b_{\gamma,1} - R_A) 
\int d^2 \vec{b}_{\gamma,2} \, \Theta (b_{\gamma,2} - R_A) \\
&& \qquad\quad \times
\int d \xi_1 d \xi_2 d x_1 d x_2
N_{\gamma\gamma} (\xi_1, \vec{b}_{\gamma,1}; \xi_2, \vec{b}_{\gamma,2})  
{\cal G}^{\rm A}_{gg} (x_1, \vec{b}_{\gamma,1} - \vec{b}; x_2, \vec{b}_{\gamma,2} - \vec{b})
\frac{d^2\hat{\sigma}_{\gamma g \rightarrow c\bar{c}}}{dy_{c}dy_{\bar{c}}}
\frac{d^2\hat{\sigma}_{\gamma g \rightarrow b\bar{b}}}{dy_{b}dy_{\bar{b}}} \,,
\nonumber
\end{eqnarray}
where $N_{\gamma\gamma}$ and ${\cal G}^{\rm A}_{gg}$ are the nuclear di-photon 
and di-gluon densities, and the impact parameters from the center of the nucleus for each of the two gluons 
are $\vec{b}_{g,i} + \vec{b}_{p,i}= \vec{b}_{\gamma,i} - \vec{b}$. Following Ref.~\cite{Huayra:2019iun}, 
we neglect possible correlations between photon and gluon exchanges that belong to separate 
SPS subprocesses. Such an approximation implies factorisation of di-photon $N^{\rm A}_{\gamma\gamma}$ 
and di-gluon ${\cal G}^{\rm A}_{gg}$ densities justified in the high-energy regime $\xi_{1,2},x_{1,2}\ll 1$ 
\cite{Blok:2011bu,Blok:2013bpa,Chang:2012nw,Gaunt:2009re,Rinaldi:2015cya},
\begin{eqnarray} \nonumber
    N_{\gamma\gamma} (\xi_1, \vec{b}_{\gamma,1}; \xi_2, \vec{b}_{\gamma,2}) & = & 
    N_\gamma (\xi_1, \vec{b}_{\gamma,1}) 
    N_\gamma (\xi_2, \vec{b}_{\gamma,2}) \,, \\
    {\cal G}^{\rm A}_{gg} (x_1, \vec{b}_{\gamma,1} - \vec{b}; x_2, \vec{b}_{\gamma,2} - \vec{b}) & = & 
    {\cal G}^{\rm A}_{g} (x_1, \vec{b}_{\gamma,1} - \vec{b}) 
    {\cal G}^{\rm A}_{g} (x_2, \vec{b}_{\gamma,2} - \vec{b})
    \label{factorisation}
\end{eqnarray}
where
\begin{eqnarray} 
{\cal G}^{\rm A}_{g} (x, \vec{b}_{\gamma} - \vec{b}) 
 & = & 
\int d^2 b_{p} \, G^{\rm A}_g (x, \vec{b}_{\gamma} - \vec{b} - \vec{b}_{p} ) \rho(\vec{b}_{p}).
\end{eqnarray}
This equation, together with Eqs.~(\ref{Gg}) and (\ref{eq:Tg}), enables one to represent the effective 
cross section that corresponds to the first contribution in Fig.~\ref{fig:two-contributions} (left) 
as follows (c.f.~Ref.~\cite{Huayra:2019iun})
\begin{align}
\sigma^{-1}_\text{eff,I}(\xi_1, \xi_2) 
\equiv \int d^2 b \, \Theta(b - 2R_A)  
T_{g\gamma}(\xi_1, b) T_{g\gamma}(\xi_2, b) \,,
\label{Eq:eff-I}
\end{align}
such that the nuclear density function enters via $T_{g\gamma}(\xi_1, b)$, as defined in Eq.~(\ref{eq:Tg}). 
In the UPC kinematics, as usual, the distance between the nuclear centres $\vec b$ (i.e.~the collision impact parameter) 
is integrated out over the region of large $b>2R_A$.

\subsection{Scattering off a single nucleon}
\label{Sect:case-II}

Since both gluons interacting with two projectile photons in the DPS mechanism as depicted in Fig.~\ref{fig:two-contributions} 
(right) belong to the same nucleon, there is no factorisation between the amplitudes for the corresponding SPS processes. In this case, 
the inverse effective cross section can be obtained in analogy to Eq.~(\ref{Eq:eff-I}) as an integral over the impact parameter of the collision $\vec b$ outside both the nuclei, i.e.
\begin{eqnarray}
\sigma^{-1}_\text{eff,II}(\xi_1, \xi_2) \equiv \int d^2 b \, \Theta(b - 2R_A)  
\int d^2 b_p\,  \rho(\vec b_p)\,  \tau_{g\gamma}(\xi_1,\vec{b}_p + \vec{b})
\tau_{g\gamma}(\xi_2,\vec{b}_p + \vec{b}) \,,
\label{Eq:eff-II}
\end{eqnarray}
where $\tau_{g\gamma}$ is the differential inverse overlap function defined as follows
\begin{eqnarray}
\tau_{g\gamma} (\xi,\vec{b}_p + \vec{b}) = \frac{1} {\overline{N}_{\gamma}(\xi)} 
\int d^2 b_{\gamma}  \, \Theta(b_{\gamma} - R_A) N_{\gamma}(\xi,\vec{b}_{\gamma}) 
f_g(\vec{b}_{\gamma} - \vec{b}_p - \vec b )
\end{eqnarray}
Indeed, this function depends on the momentum fractions of the interacting
photons $\xi_{1,2}$. Due to an additional integral over the nucleon impact parameter $\vec b_p$ w.r.t. to the center of the nucleus, such dependencies 
do not factorise, being an important example of correlations in 
the two-gluon distribution.

\subsection{Pocket formula}
\label{Sect:pocket}

In order to combine the two contributions discussed above to the resulting DPS $c\bar c b\bar b$ production cross section, 
one has to make sure that the same nucleon does not interact twice when considering the first process in Fig.~\ref{fig:two-contributions} (left). 
Besides, it is important to take into account that all the nucleons in the target nucleus participate in the interaction corresponding to 
the second process in Fig.~\ref{fig:two-contributions} (right). Requiring that these two conditions are satisfied simultaneously, 
we can construct the generalised pocket formula for the DPS cross section
\begin{align}
\frac{d^2 \sigma^\text{DPS}_{AA\rightarrow A(c\bar{c}b\bar{b})X}}{d y_c d y_b}
= \int d \xi_1 \int d \xi_2 \frac{1}{\sigma^{\rm AA}_\text{eff}(\xi_1, \xi_2)}
\frac{d^2 \sigma^{\rm SPS}_{AA\rightarrow A(c\bar{c})X}}{d y_c d \xi_1}
\frac{d^2 \sigma^{\rm SPS}_{AA\rightarrow A(b\bar{b})X}}{d y_b d \xi_2} \, .
\label{Eq:pocket}
\end{align}
Here, the differential SPS cross sections $AA\rightarrow A(Q\bar{Q})X$ for $Q=c,b$ can be found starting from Eq.~(\ref{eq:SPS}), and
effective cross section is composed of two contributions from each of the DPS processes discussed above
\begin{eqnarray}
\frac{1}{\sigma^{\rm AA}_{\rm eff} (\xi_1, \xi_2)} = \frac{w_1}{\sigma_{\rm eff,I} (\xi_1, \xi_2)} +  
\frac{w_2}{\sigma_{\rm eff,II} (\xi_1, \xi_2)} \,,
\label{eq:sigm-eff_AA}
\end{eqnarray}
where $\sigma_\text{eff,I}$ and $\sigma_\text{eff,II}$ are the effective cross sections for each process 
defined in Eqs.~(\ref{Eq:eff-I}) and (\ref{Eq:eff-II}), respectively, and  the corresponding weight factors read
\begin{eqnarray}
w_1 = \frac{A-1}{A}\,, \qquad  w_2 = \frac{1}{A}
\end{eqnarray}
in terms of the atomic mass $A$. Let us now turn to numerical results.

\section{Numerical results}
\label{Sect:Results}

In the analysis of the DPS cross section of double heavy-flavor $c\bar c b\bar b$
production in $AA$ UPCs, we present the results for lead-lead nuclei collisions
firstly, with $A = 208$. The lead nucleus radius is taken to be $R_A=5.5$ fm, and
the charm and bottom quark masses are $m_c=1.4$ GeV and $m_b=4.75$ GeV, respectively. In the case of such a heavy nucleus, more than 99\% of the contribution to this process is due to the Part-I contribution that probes gluons at different nucleons. Even for a lighter target, such as the one in a lead--carbon collision, the part-I contribution to the cross-section is still above 90\%.

\begin{figure}[!ht]
\centering
\includegraphics[width=.8\textwidth]{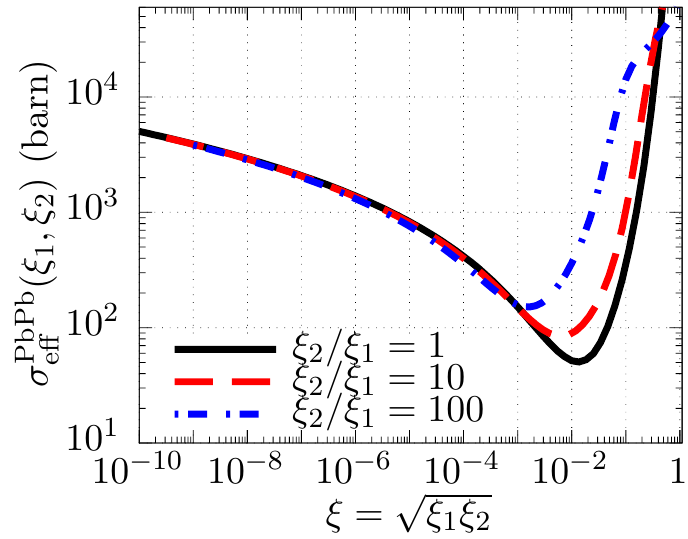}
\caption{The DPS effective cross section of $c\bar c b\bar b$ production in 
PbPb UPCs as a function of $\xi\equiv \sqrt{\xi_1 \xi_2}$.}
\label{fig:sig-eff}
\end{figure}

The behavior of effective cross section for lead-lead UPCs constructed above in Eq.~(\ref{eq:sigm-eff_AA}) 
is shown in Fig.~\ref{fig:sig-eff} as a function of the geometric mean of the two photon momentum fractions, 
$\xi=\sqrt{\xi_1\xi_2}$. This plots provides a clear idea of a typical photon energy but does not represent
the typical number densities of photons outside the given nucleus since this information has been factorised 
and absorbed into $\overline{N}_\gamma(\xi)$ function. 

Similarly to what has been observed earlier in $pA$ UPCs in Ref.~\cite{Huayra:2019iun}, in the case-I 
depicted in Fig.~\ref{fig:two-contributions} (left) 
the main contribution to the effective cross section emerges due to configurations when each of the two incoming 
photons ``meet'' the gluons inside each of the interacting nucleons. Likewise, in the case-II shown 
Fig.~\ref{fig:two-contributions} (right) the effective cross section is dominated by the configurations 
where both photons interact with gluons inside the same nucleon. Indeed, considering the symmetric 
configuration $\xi_1 = \xi_2$ as a representative example, for small $\xi$ the photons rarely overlap 
and $\sigma^{\rm AA}_{\rm eff}$ is large, while for large $\xi$ the photons are accumulated in a shell 
at the projectile nucleus periphery.

Clearly, when at some critical $\xi$ value the width of such a shell becomes 
narrower than the nucleon radius, the effective cross section grows again. Therefore, the photons do overlap for large $\xi$, but not inside the target nucleus, where most of the target gluons are located. Interestingly, in the $AA$ 
UPCs $\sigma^{\rm AA}_{\rm eff}$ develops a minimum at several times smaller than in the $pA$ UPCs 
case, i.e. at $\xi \approx 0.013$. This is due to the fact that the positions of ``active'' nucleons 
are widely distributed in the target nucleus according to the nuclear density. In the minimum of 
$\sigma^{\rm AA}_{\rm eff}$, approximately half of the photons outside the projectile nucleus have 
$b_\gamma \lesssim 2R_\text{Pb}$, i.e. located at the distance approximately equal to or smaller than
the radius of the second, target nucleus.

Note, in the considered analysis, to a good approximation, no dependence of $\sigma^{\rm AA}_{\rm eff}$ on collision energy or the factorisation scale is accounted for. In the considered formulation such an effect is subleading
as may potentially only come from an effect of broadening of the impact parameter distribution of the gluon density at high energies. As long as the gluon distribution is well localized compared to the photon distribution, it is not an important factor as the many experiments that measure $\sigma_\text{eff} \simeq 15$ mb in $pp$ collisions confirm. We leave this potentially interesting aspect for further, more quantitative, investigations elsewhere.
\begin{figure*}[!htbt]
\begin{minipage}{0.495\textwidth}
 \centerline{\includegraphics[width=1.0\textwidth]{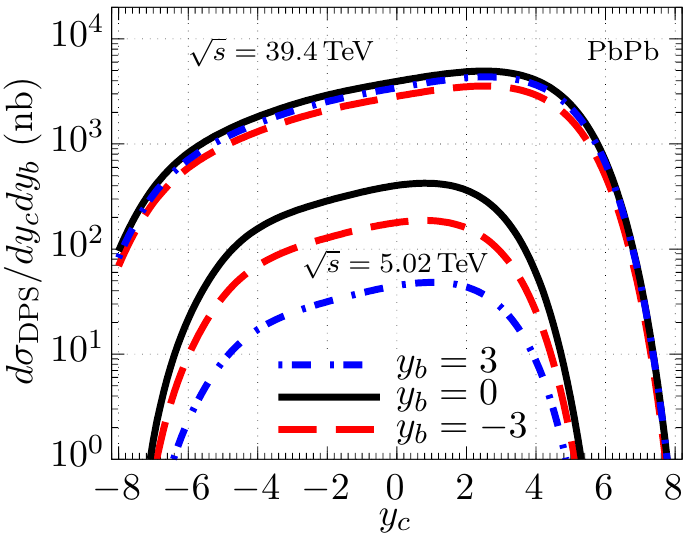}}
\end{minipage} \hfill
\begin{minipage}{0.495\textwidth}
 \centerline{\includegraphics[width=1.0\textwidth]{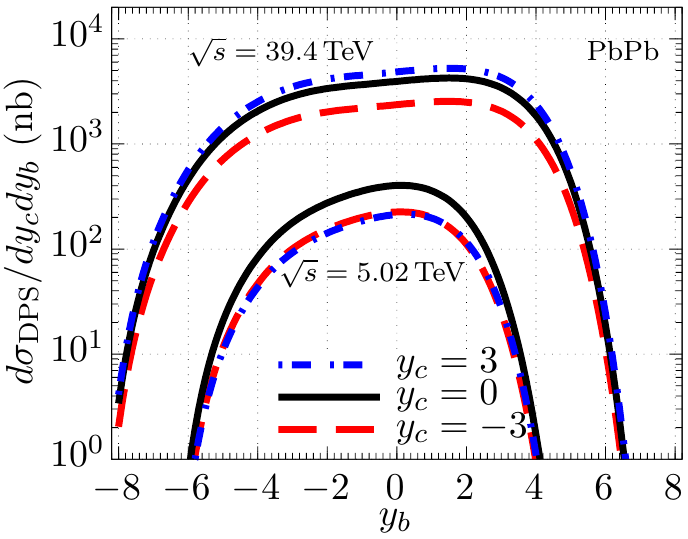}}
\end{minipage}
\caption{
The DPS $c\bar c b\bar b$ production cross section in $AA$ UPCs at typical LHC ($\sqrt{s}=5.02$ TeV) 
and FCC ($\sqrt{s}=39.4$ TeV) energies as a function of $c$-quark 
rapidity at fixed $b$-quark rapidity (left panel) and as a function of $b$-quark 
rapidity at fixed $c$-quark rapidity (right panel). }
\label{fig:DPS-ycb}
\end{figure*}
\begin{figure}[!ht]
\centering
\includegraphics[width=.8\textwidth]{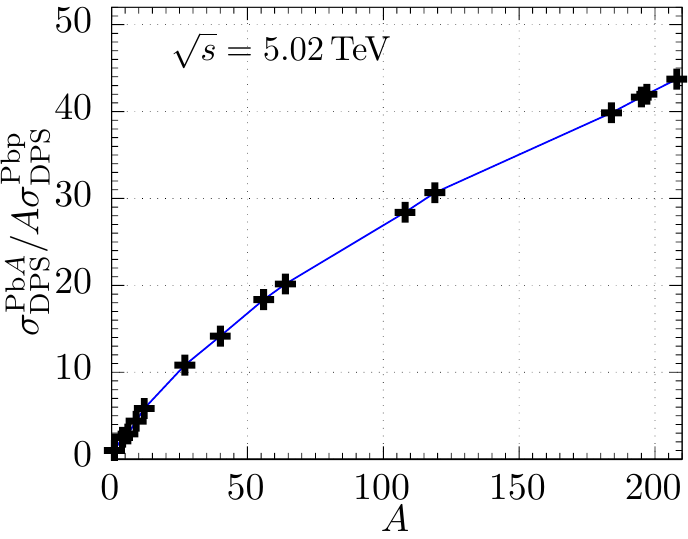}
\caption{The atomic mass dependence of the total DPS  cross section in the $AA$ UPCs 
at $\sqrt{s}=5.02$ TeV relative to that in $pA$ UPCs. Here, the projectile lead nucleus 
is adopted while $A$ of the target nucleus is a variable quantity corresponding to the  parameterisations provided by the EPPS16 nuclear PDFs.}
\label{fig:Adep}
\end{figure}

In Fig.~\ref{fig:DPS-ycb}, we present the DPS $c\bar c b\bar b$ production cross sections at two distinct 
energies, $\sqrt{s}=5.02$ TeV at the LHC and 39.4 TeV, corresponding to the planned measurements 
at the Future Circular Collider (or FCC). Here, the cross section is taken to be differential 
in $y_c$ (left panel) and $y_b$ (right panel), while integrated in $y_{\bar c}$ and $y_{\bar b}$,
respectively. As we know from earlier studies in Ref.~\cite{Huayra:2019iun}, the considered DPS 
cross sections can not be obtained by a simple rescaling of the corresponding SPS cross sections 
that makes our result nontrivial and important. Confirming this interpretation, the shape of the cross
section as a function of the first quark rapidity changes as we vary the second quark rapidity.
Additionally, since the behavior of the gluon and the photon distributions are different with $x$ 
and $\xi$, respectively, we observe that the cross section at central rapidities grows with quark
rapidity. 

Finally, in Fig.~\ref{fig:Adep} we present our prediction for the $A$ dependence of the total DPS cross 
section in the $Pb+A$ UPCs at $\sqrt{s}=5.02$ TeV relative to that in $Pb+p$ UPCs. Here, $Pb$ denotes
the projectile nucleus as a source of the di-photon flux, while $A$ corresponds to the target nucleus and 
is varied in our analysis. Similarly to the definition of the nuclear modification factor, this ratio 
is taken to be normalised to $A$. Table~\ref{Tab:total-CS-A} represents the values for the integrated 
DPS cross section (in nb) divided by target atomic number $A$ for different values of $A$.
\begin{table}{}
	\begin{center}
		\begin{tabular}{|c|c|c|c|c|c|c|c|c|c|c|c|c|c|c|c|}
\hline
A & 
1    & 4     & 6     & 9     & 12    & 27    & 40    & 56    & 64    & 108   & 119   & 184   & 195   & 197   & 208  \\ \hline
$\sigma_\text{DPS}^{\text{Pb}A}/A$ & 
1.36 & 3.42  & 3.93  & 5.93  & 7.91  & 13.70 & 19.25 & 24.95 & 27.38 & 38.59 & 41.65 & 54.14 & 56.60 & 57.05 & 59.40 \\
\hline
\end{tabular}
\vspace{0.04cm}
\caption{Integrated DPS cross section (in nb) divided by target atomic number $A$ for different values of $A$.}
\label{Tab:total-CS-A}
\end{center} 
\end{table}
We observe an approximately linear enhancement of this ratio in $A$ at large $A$, in overall consistency with simulations of the inclusive DPS cross section off proton versus nuclear targets found recently in Ref.~\cite{Fedkevych:2019ofc}. However, the growth in UPCs is steeper, since here the target gluons are probed by projectile photons that are more spread than projectile gluons and thus are more dependent on the size of target.

\section{Conclusions}
\label{Sect:Conclusions}

In this work, we present the first analysis of the double heavy-quark pair $c\bar c b \bar b$ production 
in $AA$ UPCs as a further development of our previous work \cite{Huayra:2019iun} studying such process 
in $pA$ UPCs.

We extended to UPCs the idea, already used in studies of inclusive production, of splitting the cross section 
into two parts: the first in which the photons interact with different nucleons in the nucleus (and therefore the two target gluons are not correlated) and the second in which the photons interact with two (correlated) gluons from the same nucleon. The latter is an important effect, and such correlations in the two-gluon distribution is a hot topic in the inclusive production off the proton target.

As a result, for the first time we have computed the effective cross section in $AA$ UPCs, i.e., 
two photon--two gluon double-parton scattering (DPS). As the photon impact parameter distribution 
depends on the photon energy in a non negligible way, this quantity does exhibit a certain variation 
on longitudinal momentum fractions $\xi_{1,2}$ for each incident photon. We observe an absolute minimum 
in the effective UPC cross section when $\xi_1 = \xi_2 \approx 0.013$ in the case of lead nucleus,
when the photons are mostly localised in a shell around the projectile nucleus of width roughly 
of the size of the target nucleus.

We made predictions for the PbPb DPS UPC cross section at two different energies (LHC and FCC) as a functions of both quark rapidities $y_c$  and $y_b$. The shape of the cross section as a function of the first quark rapidity changes as we vary the second quark rapidity, contrary to the case in which the effective cross section is just a constant. If the data confirm this behaviour, this would be conclusive signature of the pocket formula not having a constant $\sigma^{\rm AA}_{\rm eff}$ as indicated by our calculations.

Taking the lead nucleus as the projectile and varying the target nucleus $A$, we see a steep rise of the 
UPC DPS production cross section. This is a stronger rise than the one found in the inclusive $AA$ case. 
In short, this means that multiple parton interactions in UPCs happen much more often that in 
the inclusive case. For instance, when comparing the lead and proton targets, we obtain a factor 
of about 44 increase. This factor gives a motivation to perform such measurements at a larger 
integrated luminosity in the LHC Run 3 and, eventually, at the HL-LHC.

\section*{Acknowledgments}

This work was supported by Fapesc, INCT-FNA (464898/2014-5), and CNPq (Brazil) for EGdO and EH. This study was financed 
in part by the Coordena\c{c}\~ao de Aperfei\c{c}oamento de Pessoal de N\'ivel Superior -- Brasil (CAPES) -- Finance Code 001. 
The work has been performed in the framework of COST Action CA15213 ``Theory of hot matter and relativistic heavy-ion collisions'' (THOR). 
R.P.~is supported in part by the Swedish Research Council grants, contract numbers 621-2013-4287 and 2016-05996, 
as well as by the European Research Council (ERC) under the European Union's Horizon 2020 research and innovation programme 
(grant agreement No 668679). This work was also supported in part by the Ministry of Education, Youth and Sports of the Czech Republic, 
project LTT17018. EGdO would like to express a special thanks to the Mainz Institute for Theoretical Physics (MITP) of the Cluster 
of Excellence PRISMA+ (Project ID 39083149) for its hospitality and support.


\bibliography{bib-DPS}

\end{document}